# Micro-wires self-assembled and 3D-connected with the help of a nematic liquid crystal

Hakam Agha, Jean-Baptiste Fleury, Yves Galerne*

*Institut de Physique et Chimie des Matériaux de Strasbourg*
*UMR 7504 (CNRS–Université de Strasbourg)*
*23 rue du Lœss, 67034 Strasbourg, France*

**Abstract.** We discuss a method for producing automatic 3D connections at right places between substrates in front to one another. The idea is based on the materialization of disclination lines working as templates. The lines are first created in the nematic liquid crystal (5CB) at the very place where microwires have to be synthesized. Due to their anchoring properties, colloids dispersed into the nematic phase produce orientational distortions around them. These distortions, which may be considered as due to topological charges, result in a nematic force, able to attract the colloids towards the disclinations. Ultimately, the particles get trapped onto them, forming micro or nano-necklaces. Before being introduced in the nematic phase, the colloids are covered with an adhering and conducting polypyrrole film directly synthesized at the surface of the particles (heterogeneous polymerization). In this manner, the particles become conductive so that we may finally perform an electropolymerization of pyrrole monomers solved in 5CB, and definitely stick the whole necklace. The electric connection thus synthesized is analyzed by AFM, and its strength is checked by means of hydrodynamic tests. This wiring method could allow Moore's law to overcome the limitations that arise when down-sizing the electronic circuits to nanometer scale.

# 1 Introduction

As noticed almost half a century ago by Gordon E. Moore, the number of transistors that can be set on an integrated circuit doubles approximately every two years [1]. This down-sizing of the electronic circuits means that their overall size, their response time and power consumption are decreased too. Moreover, containing more numerous transistors, they can support higher performance softwares. Indeed, such a down-sizing is basically at the origin of the present boom in informatics. Clearly however, this law cannot go forever. Before the atomic limit, that is evident, other limitations belonging to mesoscopic physics will come. At small sizes, tunneling effects introduce electric conduction across insulators, while Coulomb blockade effects prevent conduction to occur when apparently it should. In any cases, electric currents cannot concern less than one electron. These elementary conditions clearly indicate that the down-sizing of the electronic circuits should stop much before reaching the atomic boundary.

A simple manner to overcome this difficulty suggests escaping to the third dimension, with the additional advantage of shorter interconnections. The Large Scale Integrated Circuits (LSIC) being naturally 2D, one has the possibility, in principle, to grow the circuit in the $3^{rd}$ dimension too, in a reminiscent manner as Nature connects our own brain. For this reason recently, stacks of several LSCIs have been built up, the circuits being connected by means of metal bondings with a method that is closely inspired from the classical tin soldering. Typically, the connections are realized on compressing Cu or Cu-Sn bumps at the places to connect at temperatures around 250°C – 300°C [2]. However, the typical size of these connections is relatively large ~ 15-20 μm and as far as we know, stacks of 2 pieces only have been realized up to now. Moreover, the mechanical reliability of these connections is questionable, in part because it is a rigid technology that focuses the stresses on particular hot spots, and also because of aging problems relevant to diffusions of Cu and Sn atoms, and of vacancies in the metal connections [3]. Clearly, these drawbacks are inherent to the method.

In order to overcome these difficulties in establishing 3D connections, one may think using one of the rich possibilities that liquid crystals have already demonstrated in self-assembling objects. These objects may be 1D structures [4-5], or 2D lattices [6-7]. They may be more complex, exhibiting two different scales [8]. On starting from a blue phase as the template, photonic materials have even been obtained [9]. Detailed analyses show that interactions between particles and disclination lines are often involved in such structures. The entangled disclinations work then as a rope that tightens the realized 1D [10] or 2D structures [11]. In some manner, the objects appear to be as knitted with disclination lines. However, the 1D objects are not synthesized at the right places in any of the known exemples. Therefore, we have recently proposed a radically different method that allows one to make wires that are automatically connected on the right electrodes between two substrates in front to each other [12]. The method makes use of two original properties of nematic liquid crystals, and therefore, we first fill the free space between the substrates to be connected, with a nematic liquid crystal. One of these key properties is the ability of the nematics to produce defect lines at designed places. An appropriate surface treatment is therefore applied onto the substrates. A second property of nematics allows one to materialize these defect lines and to transform them into conductive connections. Somehow, the defects work then as templates for assembling the 3D connections. In that sense, strictly speaking, the method is not a self-assembling method. Practically, we use the defect lines to attract particles initially dispersed in the nematic, and finally to form micronecklaces [13]. (In sect. 3. 3 we estimate the minimum size of the particles that may be attracted in such a way). Next, we glue the particles of the necklace together by means of an electropolymerisation process to realize the micro – and possibly nano – connections between the chosen electrodes. Clearly, the method is promising, but it is not as straightforward as it could seem at first sight. We therefore propose here to come back to the different stages of the method in order to discuss their respective difficulties. First, we notice that we need that the particles that will be attracted by the defect lines to form micronecklaces, have to be conductive in order to allow for the final electropolymerization. To achieve this property, we first cover the glass particles with a thin, conductive, and solid film of polypyrrole (PPy). This coating has moreover the advantage to provide a good substrate for the pyrrole electropolymerization that we use in the final stage to glue the particles to one another. Polypyrrole films yield planar anchoring conditions, the nematic director being parallel to the PPy surfaces. This anchoring, though less efficient than the homeotropic one [14], gives rise similarly to a nematic force that is able to attract the particles toward the defect lines, and ultimately to get them trapped onto the lines.

The paper is organized as follows. In sect. 2, we address the question of the surface treatment that is realized on the substrates in order to produce the defect lines at the places to make the electric connections. We also discuss how well are the defect lines defined with this process, and to what extent the residual uncertainty on their positions is a real and practical problem. Sect. 3 is devoted to the conditions for the particles to be trapped onto the disclination lines. The interaction energy (sect. 3. 1), the entropy difference (sect. 3. 2), and finally the free energy gain are evaluated. In particular, in sect. 3. 3 we estimate the minimum size of the particles that may be trapped on the disclinations. If they are too small, they may not be really attracted, and even if they meet accidentally the line after some Brownian motion, they may interact only weakly with it and eventually be ejected and dispersed again after some while. In sect. 4. 1 and 4. 2, we discuss the best chemical conditions to cover the particles with a conductive, thin and solid film of polypyrrole, and to realize the final electropolymerization. We then present our last materializations of the disclination lines and the tests we have performed on them (sect. 4. 3). We finally end the paper with a short discussion on the possible manners to improve the microwires again, in particular, to improve their electric conduction (sect. 4. 4) and to reduce their size towards the nano scale (sect. 5).

## 2 Defect lines

The method we propose to make automatic connections uses the defect lines of a nematic liquid crystal as templates. A first step therefore consists in producing these defect lines at the right places, i.e. between the electrodes that the connections should join, and only there. The defect lines in the nematic phase, also called disclination lines, are characterized with a singular core. They correspond to places where the orientational order of the nematic phase is broken. Practically, we obtain them on forcing a conflicting orientation on the nematic director.

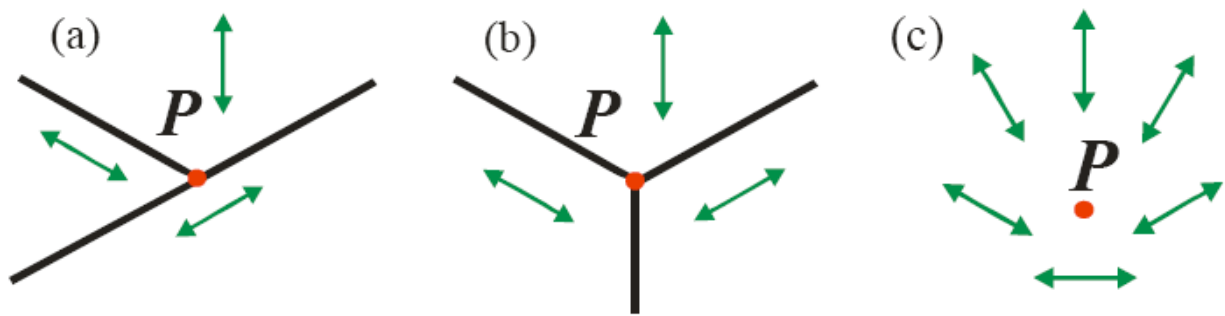

Fig. 1. Planar anchorings with conflicting orientations (green arrows). (a) Obtained with PTFE rubbings. (b and c) Realized by means of more sophisticated methods (see text).

## 2.1 Forced disclination lines

A simple manner to obtain a disclination line at the right place consists on realizing a conflicting orientation on the substrates in contact to the nematic liquid crystal. As already proposed, such a situation may be obtained by means of a planar anchoring with different azimuthal orientations in three neighboring domains forming a vertex in a definite point *P* (Fig.1 in Ref. 4). Several techniques are available to realize such an anchoring. The easiest one consists in using a rubbing technique that both applies a thin alignment layer on the substrates, and at the same time orients it along the rubbing direction. We obtained good results on rubbing the substrates with a poly(tetrafluoroethylen) (PTFE) bar at 200°C [15]. The method needs three successive rubbing paths (Fig.1a). A first one, along the vertical direction, is applied over the whole substrate while the two other paths, are applied on the lower half-surfaces of the substrate on using a half-width PTFE bar ; they are applied in the directions tilted contra-clockwise by 60° and clockwise by 60°, for the second and the last paths, respectively. This method is simple, but it yields uniformly oriented domains with their boundaries necessarily parallel to the rubbing direction. To decouple these directions from each other, one should use more sophisticated methods with appropriate masks, as oblique SiO evaporation [16, 17], or polarized UV polymerization [18], and more recently on drawing parallel grooves on the surface of a polymer with the tip of an AFM [19]. With these methods, it is possible to realize symmetric domains of uniform orientations (Fig.1b). Let us notice that such a pattern is close to, but different from those realized by Yokoyama *et al.* to obtain tristable anchorings [20]. Interestingly, the last method, namely the nano-rubbing of an oriented pattern with an AFM tip, should allow one to fabricate continuously varying orientations on changing the angle of each stripe from the previous one, in a manner reminiscent to techniques used in artistic engravings (Fig.1c). In the case that several connections have to be prepared, the rubbing method exhibits a second drawback. Then, the rubbing directions and the boundaries of the domains are not only forced to be parallel, but also, the summits $P_i$ of the domains (where the orientation is conflicting) are located on an array of parallelograms. Clearly, the other alignment techniques, recalled above, relax this condition, and the summits $P_i$ (or $Q_i$) of the domains may be placed anywhere, which could be interesting for applications.

For simplicity, let us first consider the case of only one connection between a point *P* on a substrate and a point *Q* on the other one. The two substrates receive the same surface treatment, and they are placed parallel to each other, i. e. perpendicular to the $z$-axis at a spacing distance $D$, the points ***P*** and ***Q*** being in front to each other. The free space between the plates is filled with a nematic liquid crystal – a liquid phase where the molecules are oriented along a preferred direction given by a unit vector, or director, ***n***. Practically here, we use p-pentyl-p'-cyanobiphenyl (5CB), a room-temperature nematic liquid crystal. The orientation of the nematic is then about uniform almost everywhere, i. e. in the three domains, except close to their boundaries. Along these boundaries, in the shape of capital Y, the orientation softly changes from the main orientation of a domain to the next one. The rotation is then about 60° each time and is spread over a distance of the order of the spacing distance $D$. This rotation naturally involves an elastic energy that is minimized at equilibrium, i.e. for the real orientation of the director. If one considers a path in the shape of a loop of radius larger than $D$ at the altitude $z$ (Fig.2a), we see that the director rotates by $\pi$ for every $2\pi$ rotation along the path. This shows that the loop circles a topological defect, i. e. a line along which the director is not defined, and thus where the molecules have no particular orientation. The index of the defect, often referred to as the winding number, is ±½ depending on the choice of the orientations in the three domains. The topological defect being defined independently of $z$, forms a continuous line, namely a disclination line, that joins ***P*** and ***Q***.

In the case of multiple disclination lines joining ***$P_i$*** and ***$Q_i$*** on both the substrates (see above), their respective topological indices obey a simple algebra. In particular, a loop that embraces all the disclination lines is characterised by an index that depends on the alignment conditions chosen around the boundaries of the sample and that is also the algebraic sum of all the indices of the disclination lines that the loop circles. If the nematic alignment is uniform at large distances, this index is 0, and therefore, the sum of all the topological indices of the disclination lines in the sample is 0 too.

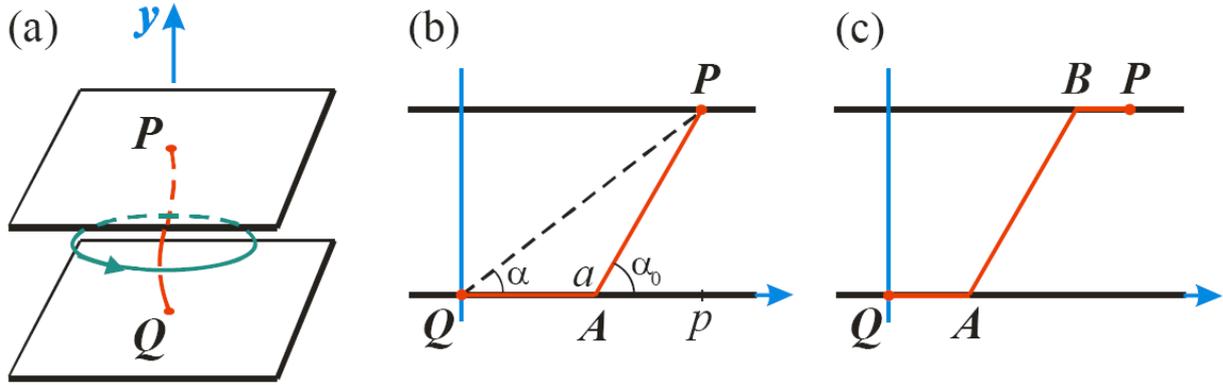

Fig. 2. Disclination joining two designed points on opposite substrates. (a) General view, **P** and **Q** being adjusted under polarizing microscope to be along the same vertical. (b and c) Side views when **P** and **Q** are not in front to each other.

## 2 . 2  Possible shapes for a forced disclination line

At this stage, one may question what is exactly the position and shape of the disclination line if **P** and **Q** are not strictly in front of each other. To discuss this point, let us call $\lambda$ the energy per unit length of the disclination line. This energy is the addition of the core energy and of the elastic energy of the whole distortion around the line, both being of the same order of magnitude [21]. When the line is stuck onto the surface, a part of its structure is virtual and its real energy is worth less, i. e. $k\lambda$ per unit length, with $0 < k < 1$. When only half of the distorted volume is rejected out of the surface of the substrate, we have $k \sim ¾$. Depending on the nature and the orientation of the disclination line referred to the substrate, and if moreover the anchoring of the director on the substrate is weak enough to break out with a low energetic cost compared to the whole line energy, the core of the line may be virtual too. In this case, we may estimate $k \sim ¼$.

In the realistic case where **P** and **Q** are not strictly in front of each other, the disclination line that joins **P** and **Q** does not necessarily lie along the straight line **PQ** at the minimum of energy. For instance, the disclination may first follow the substrate surface along **QA** before crossing the nematic layer along **AP** (Fig. 2b). Naturally, at the minimum of energy, the lengths of the line on the substrate and in the bulk are both minimal, and consequently **P**, **A** and **Q** are located in the same vertical plane. With $a$ and $p$, the abscissas of **A** and **P** referred to **Q**, being such that $0 \leq a \leq p$, the energy of the line is

$$F = k\lambda\, a + \lambda\sqrt{D^2 + (p-a)^2}\,. \tag{1}$$

Its minimum corresponds to

$$p - a = \frac{D}{\tan\alpha_0}, \tag{2}$$

where

$$\tan\alpha_0 = \sqrt{\frac{1}{k^2} - 1}\,. \tag{3}$$

Since $a \geq 0$, this implies that $\tan\alpha = \dfrac{D}{p} \leq \tan\alpha_0$, i. e. this solution (Eq. 2) is valid only if

$$\alpha \leq \alpha_0. \tag{4}$$

Thus, there are an infinity of possible disclination lines joining **P** and **Q** and composed of three parts, a bulk disclination line **AB** of slope $\tan\alpha_0$, and two surface disclination lines **QA** and **BP** that satisfy the relation QA + BP = a and that are in the same vertical plane (Fig. 2c).

In the opposite case, where $\alpha \geq \alpha_0$, the solution (Eq. 2) is impossible, and the disclination line is merely along **PQ**. The solution is then unique. This case occurs when the shift of **Q** from the vertical in **P** is less than $\dfrac{D}{\tan\alpha_0}$ (Eq. 2). Somehow, this is an estimate of the accuracy that one has to achieve when positioning the two substrates in front to each other, for realizing well-defined disclinations between them, and therefore unambiguous connections at the end of the process.

Naturally, slight departures from the equilibrium state discussed here may occur, in particular, concerning the exact position of **A** and **B**. The disclination line interacts with the solid substrate in these points, and therefore, may be trapped in a metastable state, meaning that the real points where the line anchors to the substrate could be different. For the same reason, the trajectories **QA** and **BP** of the disclination line that are adsorbed on the substrate, could slightly discard from straight lines too. Then, the general shape of the disclination line could not exactly be contained in the vertical plane. However, these departures from the ideal case should keep much smaller than *p*, so that they may be neglected in the following.

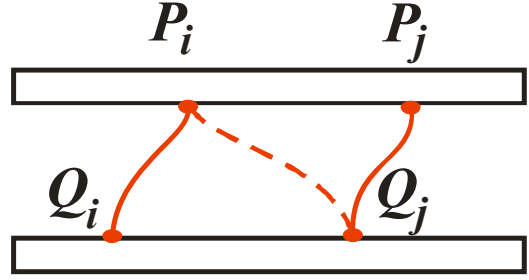

Fig. 3. Side view of self-connected disclination lines exhibiting a possible erroneous connection between two neighboring couple of points (dashed line).

### 2.3 Multiple disclination lines

When realizing several disclinations at the same time by means of conflicting surface treatments on substrates in front to each other (Fig. 3, and see also Fig. 1 in Ref. [12] for a possible surface treatment), we *a priori* have a risk that the disclinations do not connect at the right places. In order that the disclinations connect unambiguously each couple of points $\{P_i, Q_i\}$ for any values of $i$ and only them, it is necessary that they all are independent from one another. More precisely, the distortions around each disclination line should not interfere with one another. To estimate the order of magnitude of the distance where these distortions cannot interfere any more, we can make the approximation of equal Frank elastic constants. At the minimum of elastic energy, i. e. at equilibrium, the distortions obey the Laplace equation, $\Delta \boldsymbol{n} = 0$ with the condition $\boldsymbol{n}^2 = 1$ [21]. The solution to such a problem is formally difficult, but it becomes simpler if we restrict the analysis to the places where $\boldsymbol{n}$ is mostly directed along a common direction, $\xi$. Then, the condition $\boldsymbol{n}^2 = 1$ is satisfied to the first order on choosing $n_\xi = 1$, and the problem comes to solving $\Delta \boldsymbol{n} = 0$ for the two other components independently. This means that these components, perpendicular to $\xi$, are harmonic functions, i. e. they each are equivalent to a potential $V$ satisfying the equation

$$\frac{\partial^2 V}{\partial x^2} + \frac{\partial^2 V}{\partial y^2} = 0 \qquad (5)$$

with analogous boundary conditions.

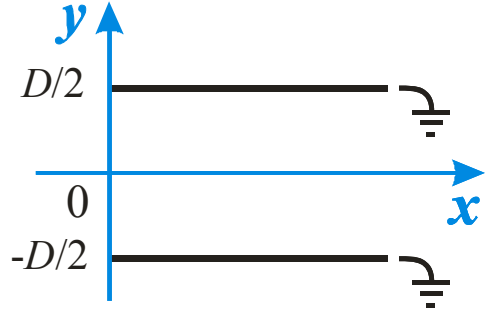

Fig. 4. Electostatic analogy for estimating the screening distance from the boundary between two rubbed domains.

Let us consider for simplicity, the vicinity of a boundary ($z$ axis) between two rubbed domains. In the x>0 region, the nematic is anchored on the plates, along the direction $\xi$. The electrostatic analogy then corresponds to parallel conductors with a constant potential, say $V = 0$ in the x>0 region (Fig. 4). Due to the boundary conditions, we may expand $V$ in Fourier series as a function of $y$. The first term in the expansion is $V = V_0(x) \cos\frac{\pi y}{D}$. Solving Eq. 5 yields the contribution of this first order term, $V(x,y) = V(0,0) \cos\frac{\pi y}{D} \exp-\frac{\pi x}{D}$. The other terms exhibiting faster variations, we deduce the screening length along the $x$ axis

$$\delta = \frac{D}{\pi}. \qquad (6)$$

Physically, this lateral screening distance results from the limited distance $D$ between the substrates. It results that as soon as a couple of points $\{P_i, Q_i\}$ are farther from other couples of points, they do not interfere. However, this condition is only a static condition. It should be complemented by a dynamic one, since the whole system may be trapped into some secondary minimum of potential as well. This difficulty may be avoided on simply orienting the transition front from the isotropic phase to the nematic one along the $y$ axis, i. e. on using a temperature gradient perpendicular to the sample. In this manner, no external action will perturb the screening along the $x$ axis, and the disclination lines should join each couple of points $\{P_i, Q_i\}$ without any error.

# 3 Condensation of a particle onto a disclination line

Impurities are known for a long time to condensate onto the defects of crystalline materials [22]. Similarly, particles with a definite anchoring (e. g. the director is supposed to be anchored perpendicular to the surface) and immersed in a nematic, are attracted toward disclination lines, and they eventually stick onto them [13]. Such a behavior is somehow akin to the observed gathering of oil droplets in areas of large director gradients [23]. The assembling occurs when the free energy of the whole system $F$ decreases, i. e. when the free energy difference $\Delta F = \Delta U - T \Delta S$ between the dispersed and the condensed states is negative, $U$, $T$ and $S$ being the potential, temperature and entropy, respectively. Let us estimate $U$ and $S$ in both states.

## 3 . 1 Interaction energy of a particle with a disclination line

The variation of the interaction energy of a particle $\Delta U$ (= $U_{cond}$ - $U_{disp}$) between the dispersed and the condensed states depends on the anchoring conditions on the particle. The case of the homeotropic anchoring onto the particle is the simplest one. It may be observed on using silica particles treated with silane [13].

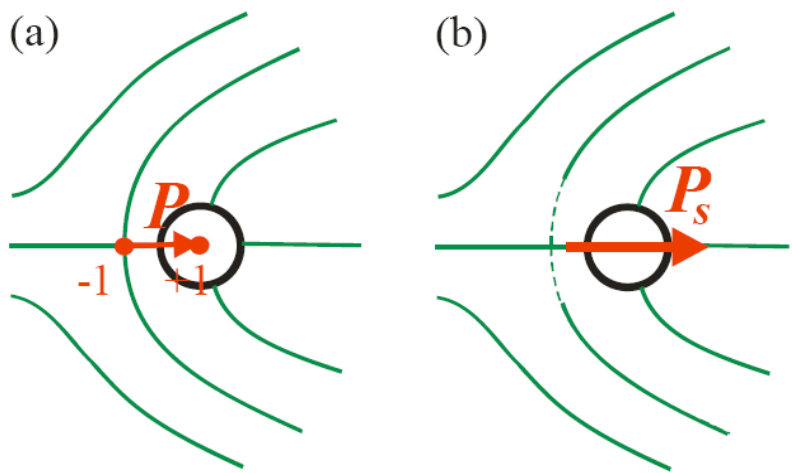

Fig. 5. (a) Dipole moment ***P*** from the point of view of the topological charges, -1 and +1, of the hyperbolic hedge-hog defect and in the centre of the particle, respectively. (b) Splay moment ***P**$_S$* of the couple particle – companion defect, generating a splay field around the particle as may be seen on the general shape of the director lines.

### 3 . 1 . 1 Homeotropic anchoring

The difference $\Delta U$ between the interaction energies of a particle with a disclination line, in the dispersed and in the condensed states, respectively, may be estimated rather simply in the case of homeotropic anchoring. We start from the remark that, when immersed in the nematic, the particle essentially introduces a splay distortion that may be considered as a vectorial charge of splay, namely a splay moment, **Ps** (Fig.5). To the lowest order, the coupling energy of such a splay moment to the splay field $\mathbf{S} = \mathbf{n}\, \nabla.\mathbf{n}$ may be written as

$$\delta U = - K\, \mathbf{Ps}.\, \mathbf{S}\ . \tag{7}$$

This term is analogous to the dipolar term in the multipolar expansion of the coupling energy between a charged line and an electric dipole moment, **S** being equivalent to the electric field produced by the disclination line, and the splay moment **Ps** , to the dipole electric moment in interaction. For symmetry reasons, **Ps** is proportional to the topological moment **P** of the system constituted of the particle and its companion defect [24]. At the minimum of energy, **Ps** orients parallel to **n** like a compass in a magnetic field. Eq. 7 then reduces to $\delta U = - K\, Ps\, \nabla.\mathbf{n}$ . Far from the line, **n** is almost uniform and thus, $\delta U_{\text{disp}} = 0$. So, $\Delta U = \delta U_{\text{cond}}$ . When a particle with an homeotropic anchoring is condensed on a disclination line, its companion defect is in fact in contact to the line, and therefore, the distance of the particle centre to the line is $\sim R$ . In order to estimate $\Delta U$ , we may now use dimensional analyses. $Ps$ being of the dimension of a surface associated to the particle, we deduce $Ps \sim R^2$, and $S = \nabla.\mathbf{n} \sim 1/R$ , since $R$ is the effective distance where to evaluate the splay distortion $\nabla.\mathbf{n}$ [24]. We may evaluate the average Frank elastic constant by means of a dimensional analysis too, on noticing that $K$ is the ratio of a typical coupling energy over a characteristic length. The coupling energy between the liquid crystal molecules is of the van der Waals type, $\sim kT$ , $k$ being the Boltzmann constant, $T$ the temperature, and $l$ the effective length of the molecules [21]. Here, 5CB being a dimer, we may estimate $l \sim$ 3 nm. So, we obtain $K \sim kT / l$ , i. e. $K \sim 13$ pN that compares not too bad with the experimental values $K \sim 5 - 10$ pN depending on temperature [25]. Finally the difference between the interaction energy of a particle $\Delta U$ (= $U_{\text{cond}}$ - $U_{\text{disp}}$) in the dispersed and in the condensed states, is

$$\Delta U \sim - kT\, \frac{R}{l}\ . \tag{8}$$

Other methods may be used to estimate $\Delta U$. One may for instance consider the additional elastic energy of the distortion produced by the particle immersed in the nematic phase, far from the disclination line and close to it, respectively. Far from the line, in the absence of the particle, the nematic is uniformly oriented, and its elastic energy is close to zero. With the particle and its companion defect, we essentially add a splay distortion that extends around at a distance $\sim R$ (Fig. 5b). The energy of the distortion is $U_{disp} \sim \frac{1}{2} K \boldsymbol{S}^2 V$, where $V \sim R^3$, and $S \sim 1/R$. Naturally, in these crude estimates, the numerical coefficients have been dropped down systematically, as has been forgotten the bend and twist distortions that are produced too around the particle. Similarly, we can evaluate the elastic energy of the system in the condensed state $U_{cond} \sim \frac{1}{2} K (\boldsymbol{S} - \boldsymbol{S'})^2 V$, where $\boldsymbol{S'}$ is the average splay field of the disclination line at the place where the particle stabilizes. We may therefore anticipate that the particle adjusts its position, and therefore $\boldsymbol{S'}$, so that the difference $\boldsymbol{S} - \boldsymbol{S'}$ is small at equilibrium compared to $\boldsymbol{S}$, and that finally $U_{cond} \sim 0$. [More exactly, the in-plane splay contributions only are considered here, as the out-of-plane splay distortions do not interfere with the 2D distortion of the line, $\boldsymbol{S'}$]. This leads again to Eq. 8. However, this approach allows us to discuss qualitatively what happens in the case of a weak anchoring of the director onto the surface of the particle. If the anchoring constant $A$ is not too weak, the distortion around the particle is equivalent to the one produces by a particle with a strong anchoring, but with a reduced effective radius $R - L$, where $L = K/A$ is the extrapolation length [21]. If the anchoring is very weak, it may eventually break down along areas on the particle surface where the anchoring torque exceeds a critical value [25], thus reducing again the distortion around the particle, and consequently its coupling energy. This case will be discussed elsewhere [14]. The above estimates may seem rather crude. For a more accurate determination of the interaction energy of a particle $\Delta U$, one would need to make numerical simulations of the distortions and of their energies in the dispersed and condensed states, as there is no analytic solution to this problem.

It is also interesting to compare the coupling energy of a particle to a disclination line (Eq. 8), which may be understood as the adhesion energy of the particle onto a disclination line, with the adhesion energy of the same particle onto a substrate. Naturally, this interaction energy is proportional to the surface of the particle in contact to the substrate $\sim R^2$, and to the van der Waals interaction between the particle and the substrate $- kT$. So, as discussed in Ref [26], the interaction energy between the particle and a substrate may be estimated as $\Delta U \sim -kT\left(\frac{R}{l}\right)^2$.

This is a marked difference from Eq. 8 that essentially arises from the different dimensionalities between the two problems, 2D here, instead of 1D in the case of Eq. 8.

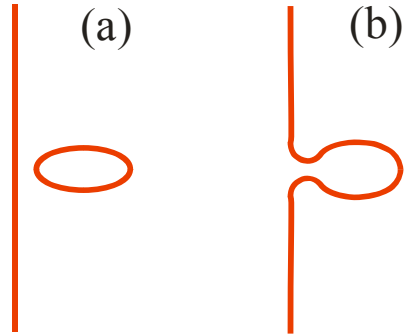

Fig. 6. Saturn ring – disclination line system. The particle itself (virtually in the centre of the ring) is not shown here. (a) Before and (b) after they get connected.

In a third approach, we may discuss the interaction between the particle and the line in terms of the interactions between the respective defects. We notice first that provided some energy is given to the particle, its companion defect may open up to give rise to a so-called Saturn ring [24]. Then, this ring-shaped disclination may combine to the disclination line. It thus reduces the overall disclination length by ~ 2R (Fig. 6), and saves an energy ~ 2R λ. Subtracting the initial energy necessary to form the Saturn ring, we deduce the interaction energy of the particle with the disclination line $\Delta U \sim R \lambda$. A simple dimensional analysis yields $\lambda \sim K$, which leads to Eq. 8 again. Let us finally notice that Eq. 8 is consistent to within a factor of 5 only with the experimental data [27].

### 3.1.2 Planar anchoring

In fact, we coat the particles with a PPy film in order to make them conductive before being dispersed in the nematic. The PPy coating also gives them planar anchoring properties, i. e. with the easy axis parallel to the surface. This anchoring is not as simple as the homeotropic case discussed above since, if the zenithal angle is fixed at 90° referred to the normal, the azimuthal angle keeps undetermined. Complete anchoring conditions should thus specify too the exact direction of the easy axis over all the surface of the particle, with at least, two vertices. When immersed in the nematic, the vertices will produce two surface point defects, or boojums, on the particles. However, in the case of extremely weak azimuthal anchoring, the problem significantly simplifies. The azimuthal angle adjusts then everywhere to minimize the overall energy. Since the boojums, associated to the vertices, exhibit splays of same sign, they repel each other, and they finally stabilize in diametric positions, consistent with the symmetry of the particle (Fig. 7a). Clearly, the nematic distortion that such a particle produces all around is the combination of 2 opposite splay moments, that both together correspond to a splay quadrupole, $\boldsymbol{Qs}$ . The coupling energy of the particle with a splay field may then be obtained on differentiating Eq.7. This yields $\delta U = - K \, \boldsymbol{Qs} : \nabla \boldsymbol{S}$ [14]. Taking into account that the quadrupole orients itself along the director $\boldsymbol{n}$ as the dipole moment in the homeotropic case, we have $\boldsymbol{Qs} = Ps \, R \, \boldsymbol{n} \, \boldsymbol{n}$ , and the coupling energy becomes :

$$\delta U = - K \, Qs \, \boldsymbol{n} \cdot \nabla ( \nabla \cdot \boldsymbol{n} ) . \qquad (9)$$

Such weak azimuthal anchorings occur when appropriate polymer coatings are used [29-30], or contrarily to what is often reported, they are also observed on extremely clean glass substrates [31]. Except there, the exact field of the easy axis onto a particle is generally more complicate than exposed above, the 2 vertices not being in diametric positions anymore. However, the problem may be simplified on discarding the regions close to the particle and to the disclination line where the distortions essentially contribute to the proper energy of the particle and of the disclination, respectively. In the intermediate region, the distortion is smoother and its energy may be estimated on considering, as in sect. 2. 3, that the director $\boldsymbol{n}$ does not discard much from a general direction. Then, the perpendicular components of $\boldsymbol{n}$ are harmonic functions and they may be expanded in multipolar series. (A similar approach as in sect. 3. 1. 1 may be used to reach to the same conclusion). So, the first term is dipolar, and though in principle, it should be the superposition of splay, bend and twist distortions, it is essentially of bend nature here. This term arises from the coupling of the bend moment produced by the particle $\boldsymbol{Pb}$ to the bend field of the disclination $\boldsymbol{B} = \boldsymbol{n} \wedge \mathbf{curl} \, \boldsymbol{n}$ :

$$\delta U = - K \, \boldsymbol{Pb} . \boldsymbol{B} . \qquad (10)$$

This bend coupling energy is the analogue of Eq. 7 for the splay distortion. With the same arguments as in Sect. 3 . 1 . 1 (on just rotating the director by 90°), we may estimate $B \sim 1/R$ and

$Pb \sim R^2 \times \theta$, where $\theta$ is the angle between the radii that join the two vertices (Fig. 7b). This angle essentially depends on the anchoring of the director on the particle. It cannot be controlled experimentally. Practically, it may be different on each particle. Naturally, if $\theta$ is small, this bend term becomes negligible, compared to the next term in the splay expansion, i.e. to the splay quadrupolar term (Eq. 9). Let us however notice that the exact crossover between the two regimes depends on the distance particle-disclination since the dipolar energy decreases as $1/r$ while the quadrupolar one varies as $1/r^2$.

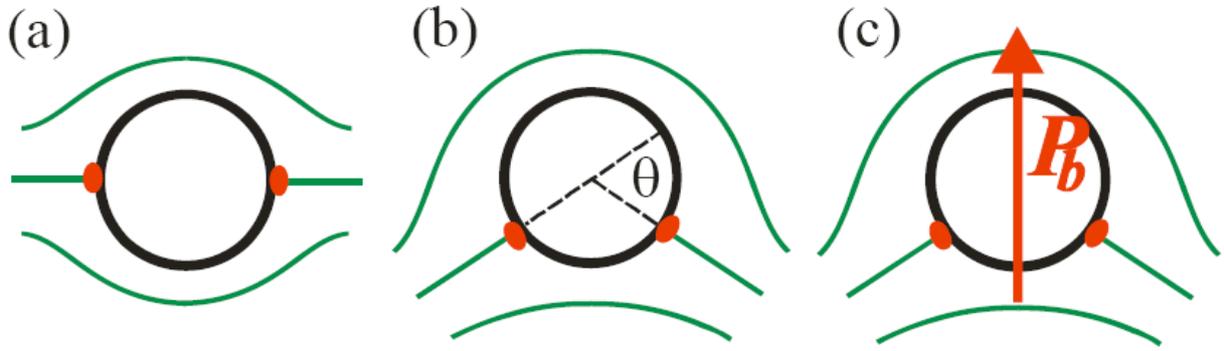

Fig. 7. Particle with planar anchoring. (a) Boojums in diametric positions. (b) Strong azimuthal anchoring. The boojums keep blocked at an angle $\theta$. (c) Resulting bend moment **Pb** carried by the particle.

## 3.2 Entropy difference

In order to obtain the free energy difference between the dispersed and the condensed states, we need now to calculate the entropy difference between the two states. The entropy is given by the Boltzmann formula $S = k \ln \Omega$, where $\Omega$ stands for the number of accessible states. To within a multiplicative constant, $\Omega$ is given by the ratio of the accessible volume for the particle over its own volume, $V/v$. Therefore, we have $\Omega_{cond} \sim D/2R$ and $\Omega_{disp} \sim \frac{3}{4\pi}\left(\frac{D}{R}\right)^3$ in the condensed and the dispersed states, respectively. In the dispersed state, we have considered that the accessible volume was laterally bound by the screening length δ that we take here ~ $D/2$. We deduce $\Delta S$ (= $S_{\text{cond}} - S_{\text{disp}}$) ~ $k \ln 2 \left(\frac{R}{D}\right)^2$ and therefore $\frac{\Delta F}{kT} = -\frac{R}{l} - \ln 2 \left(\frac{R}{D}\right)^2$.

## 3.3 Trapping condition

So, the particles will be trapped on the disclination line if $\Delta F < 0$, i. e. if $\ln \frac{1}{2}\left(\frac{D}{R}\right)^2 < \frac{R}{l}$. With $D/R$ ~ 100 or so, the logarithm is worth roughly 10, but is relatively insensitive to the exact value of $D/R$, so the trapping condition becomes,

$$\frac{R}{l} > 10, \qquad (11)$$

which means that the particle radius should be larger than ~ 30 nm for the particles to be definitely trapped on the disclination line. Naturally, in order to get a more precise value of this threshold, numerical simulation would be necessary. Nevertheless, we see that this condition is a little bit more restrictive than the condition $\frac{R}{l} > 3$ necessary for the particles to be trapped on a substrate [27]. This estimate also explains why we failed when we tried to use monomers instead of particles. In this earlier project, the disclination lines were supposed to confine the monomers as in a micro-reactor, and to allow for a direct polymerization of the wire at their places. So, in the work discussed here, we make use of particles of radius ranging from 1 μm to 40 nm.

## 3.4 Necklace of particles

Interestingly, the attraction of the disclination line does not depend on the number of particles already stuck on it, since the nematic force is mediated by the distortion around the line and that this distortion is not significantly changed by trapped particles at distances much larger than the particle radius. Similarly, the trapping power of the line is not modified too by the presence of particles stuck on it, because both the interaction energy and the entropy calculations keep roughly the same as in the case of a bare line (sect. 3.1 and 3.2). So, the disclinations eventually attract and stick a large number of particles and, after an iterative process, we obtain a kind of micro or nano necklace. On gathering the particles one-by-one after some Brownian motion, we finally get a rather uniform necklace. Sometimes, aggregates may be observed on the necklace. They often form before reaching the disclination line [probably due to an insufficient ultrasound treatment when dispersing the particles in the nematic] and they are attracted toward it as larger particles.

Let us notice that, because of gravity, the density of the particles that are dispersed in the nematic liquid crystal is not uniform. They are more numerous close to the lower plate than to the upper one, and this inhomogeneity is reflected on the particle density along the necklace, more particles being stuck in the bottom part of the necklace, forming small aggregates. According to Boltzmann law, the particle density is proportional to $exp - \frac{\delta m\, g\, z}{kT}$, where $z$ is the altitude, $g$ is the acceleration of Earth gravity, $\delta m = \frac{4\pi}{3} R^3 \delta\rho$, and $\delta\rho = \rho_{silica} - \rho_{5CB} \sim 1000$ kg/m$^3$. However, if $\frac{\delta m\, g\, D}{kT} < 1$, the exponential is roughly a constant and the particle inhomogeneity may be neglected. This is the case for $R < \left[ \frac{3}{4\pi} \frac{kT}{\delta\rho\, g\, D} \right]^{1/3}$, i. e. for $R < 100$ nm, since to make the observation easier, we chose a large spacing distance, $D \sim 150$ μm, between the plates.

## 4  Materialization of a disclination

Before being dispersed in the nematic liquid crystal, the silica particles are covered with a PPy film that gives them both the properties of electric conduction and of planar anchoring. The nematic director is then anchored parallel to the surface as we have independently verified on specific flat samples between crossed polarisers. This anchoring is at the origin of the interaction between the particles and the disclination, and therefore is necessary to assemble the necklace. By means of an electropolymerization of pyrrole monomers dissolved in the nematic, we finally achieve the materialization of the disclination line. In the next two sections, we examine the conditions of these polymerization processes.

### 4 . 1  Polypyrrole coating of particles

In order to coat the particles with a tough and adherent polymer film, we synthesize PPy directly on the solid surfaces in the surface polymerization regime. At higher concentrations, i. e. above a threshold, a volume polymerization takes place too and becomes dominant. PPy aggregates are then essentially synthesized in the bulk. They diffuse and settle onto the substrates after a while, but the film that they produce keeps labile contrary to the film obtained by surface polymerization. Typically, with [Py] = 0.036 mol.L$^{-1}$, [FeCl$_3$] = 0.0092 mol.L$^{-1}$ and a reaction time of 1 hour, we obtain PPy films about 50 nm thick and exhibiting conductivities of about 40 S.m$^{-1}$, all properties appropriate for coating particles to be used as elementary bricks in the conducting microwires [32].

## 4 . 2 Electropolymerization of the necklace

After the particles have been coated with a PPy film, they are dispersed in 5CB. This is possible if the silica particles repulse one another by means of electric charges on their surface. We therefore determined the electric charge carried by the particles when dispersed in 5CB, on measuring the fall velocity in the presence of an electric field applied upwards. Typically, 100nm radius bead are measured to bear ~ 5e when just dispersed in the nematic, which roughly corresponds to about 40 electrons per µm². However, this electric charge gradually decreases with time and typically after one day, the particles lose about half their charge. This shows that the radius of the particles cannot be decreased much below 100nm. Otherwise, the particles that have lost their electric charge could form aggregates. In practice, we do not use particles of radius smaller than 40 nm, a limitation that we have already encountered above for entropy reasons (sect. 3. 3). However, this limitation concerns spherical objects only. One could overcome it on using elongated particles as Carbon Nano Tubes (CNT) of similar surface and therefore of similar electric charge [32].

So, a disclination line being prepared at the place where we intend to synthesize a microwire (sect. 3), particles are trapped onto it until they form a continuous micro or nano-necklace. When the line is saturated with particles, its shape is not flexible any more as are the disclinations in nematics, but exhibits zigzags that are characteristic of a non-fluid behavior. This indicates that the particles are in solid contact to one another (Fig.8). We then perform a second polymerization in order to fix them definitely to one another. A small amount (0.02 wt%) of pyrrole monomers is therefore initially solved in the 5CB nematic together with some (0.005 wt%) AgBF$_4$. This salt plays the previous role of FeCl$_3$, Ag$^+$ being the oxidant, and BF$_4^-$ the doping ion, but exhibits a larger solubility in 5CB. Except for this minor difference, the chemical reaction is essentially the same as for the initial coating of the particles [33]. Practically, the concentrations chosen here are low enough for the synthesis keeps below the volume polymerization threshold, thus preventing bulk gelification to occur.

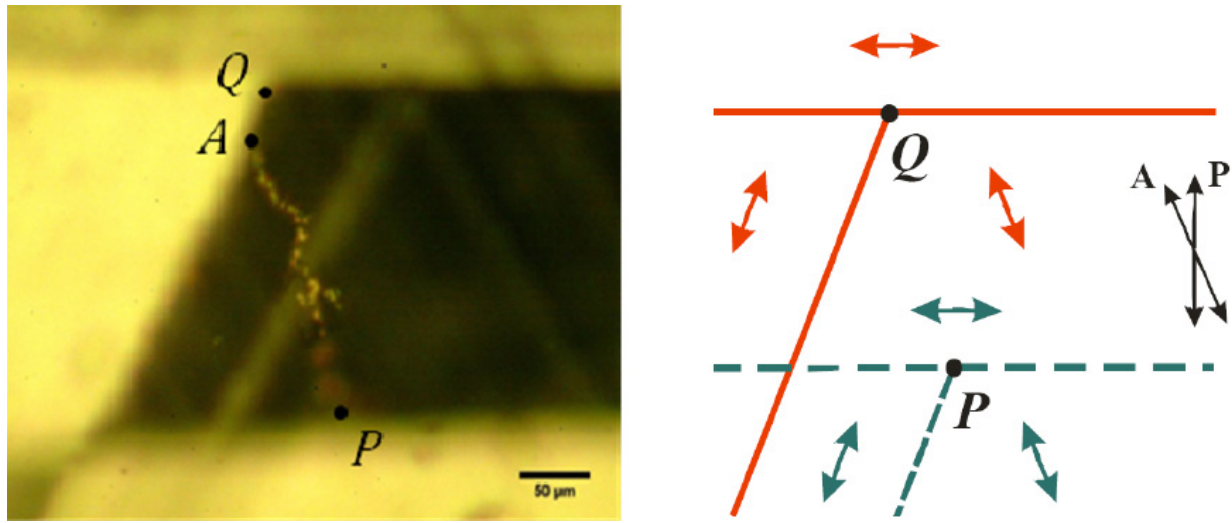

Fig. 8. Photograph of silica beads trapped onto a disclination line, observed in black field between polarizers (left) and schematic of the sample (right), showing the anchoring directions (cf. Fig. 1a) onto the upper substrate (red arrows) and lower substrate (green arrows). The central region is twisted, and due to a wave-guiding effect, is extinguished when the polarizers are oriented as indicated by the black arrows. The plates are off-centered from each other to make the necklace more visible. The necklace is connected to the electrodes at the points *P* and *A* (see *sect 2. 2*), on the lower and upper substrate respectively. The focus is on the upper middle of the 3D necklace, the lower part (around *P*) being blurry. The bar is 50µm long.

Though the $Ag^+$ ion is able to oxidize the pyrrole monomer by itself, we aim now to trigger the polymerization at the right moment, and to localize it only between each particle of the necklace in order to glue them together cleanly. To do so, we perform an electropolymerization of the necklace on applying a small alternating voltage (750 mV) at a low frequency (300 mHz) between both the electrodes for 6 hours [32]. In this manner, each neighboring particle works as a microelectrode as soon as Brownian motion opens up a gap between them. The oxidation is then realized electrolytically. Each particle in the necklace alternately plays the role of an anode, captures electrons and therefore oxidizes the adsorbed pyrrole monomers in radicals. They further react and produce PPy at the right places provided that they are not dispersed away, e. g. by diffusion [34]. In order to minimize the hydrodynamic current that could also carry pyrrole radicals away, and consequently disperse the PPy materials around, the voltage is reduced to the lowest level compatible with electropolymerization. Let us also notice that, for each half-period, one particle over two is a cathode that reduces the $Ag^+$ ions and produces nodules of silver metal. As the role of the electrodes is reverted after each electric alternation, a cement of PPy incrusted with silver nodules is finally synthesized between the particles. They indeed build micro-bridges that finally fix the necklace into a micro-chain of particles.

We have inspected the quality of this micro-chain of particles on performing AFM observations. To this aim, we break the chain and project it on one of the plates. We first remove the liquid crystal from the sample and clean it with several passages of ethanol to make sure that no liquid crystal is left on the surface. The sample is then left for the ethanol to evaporate before performing AFM observations. We observe that though the connections between the particles are generally satisfactory, they are sometimes extremely tiny (Fig.9).

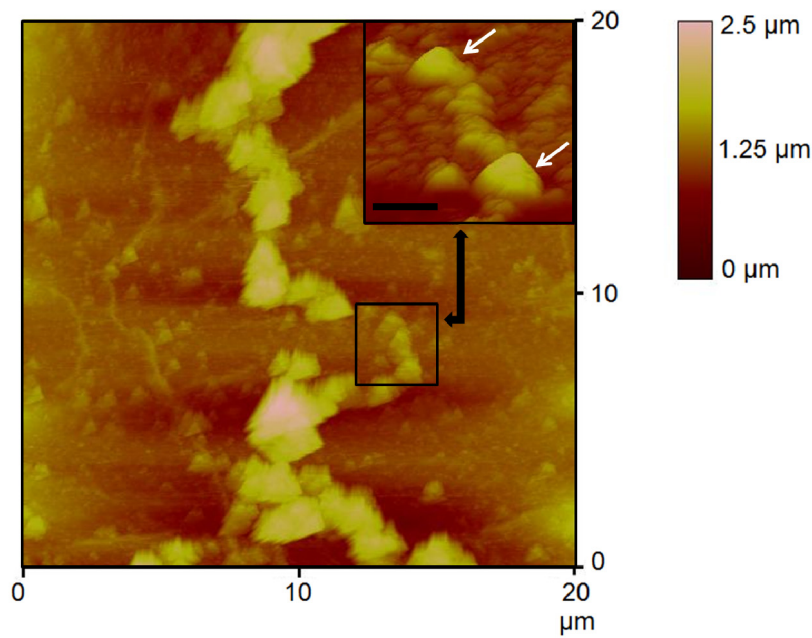

Fig.9. AFM image of a chain of silica beads projected on the lower substrate after breaking. The diameter of the beads is 700 nm including their PPy coating (see sect. 4. 1). The triangle-like patterning arises from the convolution of the pyramidal tip with the beads in the chain. The color scale gives the height of the chain in different places. The inset shows an enlarged 3D view of a polymer junction between two beads (white arrows). The bar is 1µm long.

An explanation for such a difference in the junctions between the particles could be that some of them exhibit a good conductance even when tiny. Then, the particles that they join keep at the same potential so that they do not play the role of microelectrodes any more. The synthesis of PPy is then stopped at this place. In this scheme, we could observe some decoupling between the mechanical and the electrical properties in a chain of particles. A particular junction could exhibit good mechanical properties with a poor conductance, and reciprocally. A remedy to this problem could be to synthesize the PPy in the junctions with more uniform electric values. However, we have not succeeded yet to reduce these irregularities.

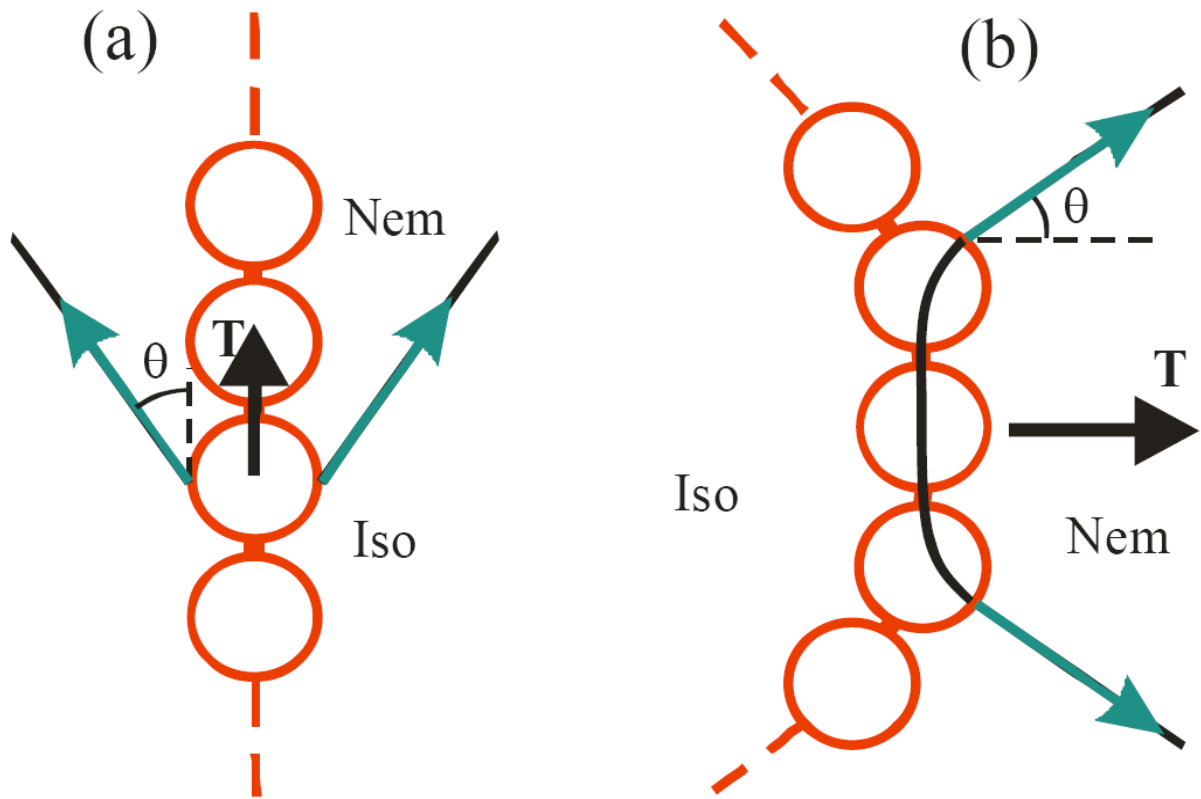

Fig. 10. The capillary forces (green) that arise at the NI interface produce a mechanical tension ***T*** on the chain of particles bounded by PPy bridges. (PPy is in red). (a) The average direction of the NI interface (black) is perpendicular to the chain, itself being roughly perpendicular to the substrates (not shown here). (b) Parallel to the chain.

## 4.3 Evidence of the microwire

After the electropolymerization is performed, we verify that the microchain is continuously connected. To do so, we first suppress the nematic force that gathered and hold the particles onto the disclination line, by heating the liquid crystal into the isotropic phase. During this process the microchain has to be strong enough to withstand the capillary forces that arise at the nematic-isotropic (NI) interface. Two extreme situations may be encountered then. In the first one (Fig.10a), the NI interface is roughly perpendicular to the chain apart for a meniscus effect. A mechanical force ***T*** ~ $2\pi R \sigma \cos\theta$ pulls on the chain, $\sigma$ being the surface tension of the NI interface, and $\theta$ the angle of the interface with the chain direction, e.g. with the perpendicular to the substrates. In the opposite configuration (Fig.10b), the NI interface and the chains are roughly parallel. Then, if the chain is loose as on the figure, the force that pulls it may be roughly estimated to be *N* times larger than in the previous case, *N* being the number of particles in contact to the interface. Clearly, this situation is less favorable. We therefore always work in the case of Fig.10a to preserve the integrity of the wire.

Once the transition is done, we obtain a 3 dimensional chain connected from the upper substrate to the lower one (Fig.11). But this test is not sufficient since we cannot *a priori* discard the possibility that the entire sample is polymerized. So, in order to prove that this is not the case, we verify that the whole sample keeps able to flow in the isotropic phase while preserving the microchain.

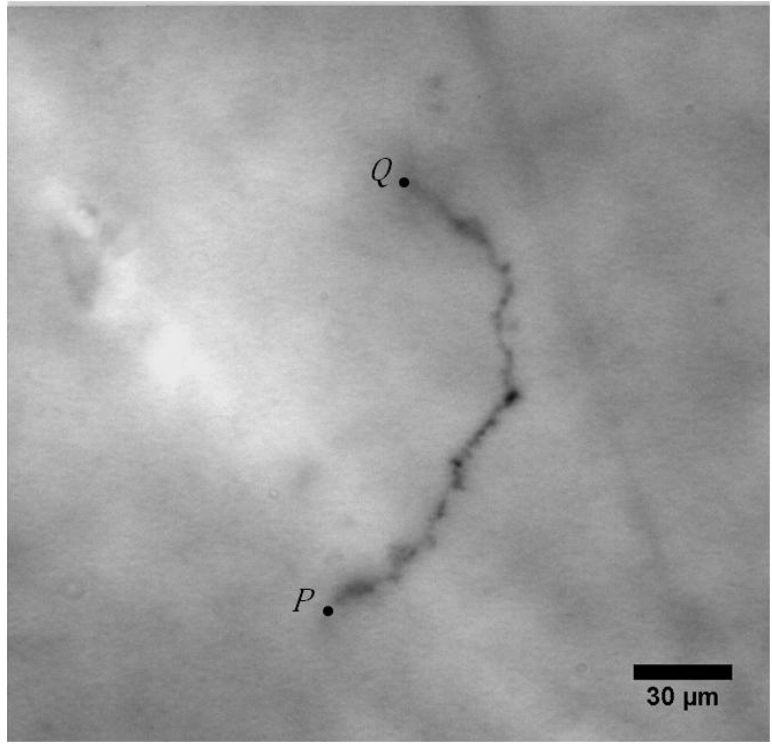

Fig. 11. Optical microscope image of the middle part of a 3D wire in the isotropic phase. ***P&Q*** are the contact points of the wire on the substrates (see *sect 2. 2*).

So, we perform a micro-flow of the isotropic liquid crystal around, to verify that the whole system is not completely polymerized. In the attached video [35], is shown a 3D wire undergoing a flow of 25 $\mu m.s^{-1}$ velocity, and withstanding the tension the flow exerts on it. The drag force $f$ per unit length of the wire, supposed to be a cylinder of radius $R$ and length roughly equal to the spacing distance $D$ (= 140 μm), can be estimated by $f \sim 4\pi\eta\upsilon/[\ln(D/2R) – 0.9]$, where $\upsilon$ is the velocity of the flow, and $\eta$ is the shear viscosity of 5CB in the isotropic phase ($17 \times 10^{-3}$ Pa.s) [36, 37]. Taking into consideration the dip of the wire to be of the order of $L/10$, we can estimate the tension to which the wire can resist to be larger than $T \sim 10\,fL = 15$ nN. This value is well improved compared to our first synthesis [12]. It is now above the rupture limit known for C-C chemical bonds ~ 5 nN [38]. This however does not mean that the connections between the particles in a chain are due to covalent bonds between pyrrole monomers. They could be due as well to a large number of van der Waals interactions between polymer chains, each exhibiting a typical strength at rupture ~ 0.1 nN. Let us also notice that this order of magnitude for the wire rupture keeps insufficient to withstand without damage the capillary forces at the surface of the liquid, and therefore to allow one to remove the nematic from the spacing between the substrates, or to change it by another liquid.

In fact, the rupture point is generally located on the substrate itself, and more precisely on the upper one since the particle density is lower there. Clearly, the PTFE treatment does not favor the binding of the wire onto the substrate. To help it, we scratched the substrates with a needle around *P* and *Q*. Another means could be to use a different planar anchoring treatment as polyvinyl alcohol (PVA) instead of PTFE. One could also reinforce the adhesion of the wire to the electrodes on extending the segments *PB* and *QA* of adhesion of the chain on the substrates (Fig. 2c).

## 4 . 4  Electric resistance

Let us notice that the wire conducts electricity because its silica particles are covered with a thin layer of conductive polypyrrole and that they are finally stuck together by a small joint of polypyrrole again, with polymerization conditions leading to a conductivity of ~ 40 S.m$^{-1}$ (Sect. 4. 1 and 4. 2). Another argument for the conduction properties of the wire is that basically it is obtained by means of electropolymerization, which process needs an electric current to pass through.

However, we could not completely remove the liquid crystal from the space between the electrodes, and this liquid crystal, containing essentially $Ag^+$ and $BF_4^-$ ions, inevitably by-passes the wire. We therefore could not measure its resistance directly. Most probably, this resistance is too large for immediate applications. The next step could then be to reduce the wire resistance by electroplating it with an appropriate thickness of silver or gold. In order to realize this electroplating, we have to replace the nematic liquid crystal by an appropriate electrolyte in the interval between the substrates. This manipulation is difficult since, as mentioned in the previous section, the polymerized chain of particles is not strong enough to withstand the capillary forces between these two liquids. To avoid this problem, we intend to use an intermediate liquid as alcohol, that is a good solvent of both the liquids, 5CB and water, and that consequently, would not produce any interface between them, nor capillary forces during the drainage.

## 5 Conclusions

In summary, we have described in the details how automatic 3D connections can be realized. The process is essentially based on two exceptional properties of nematic liquid crystals. The first property we use is the ability of nematics to form disclination lines in designed locations. The second special property is the existence of the nematic force itself that is able to drive small objects toward disclination lines. This last step makes the prelude to a materialization of disclination lines. However, the objects to be used in this materialization, have to be large enough to get dispersed in the nematic liquid crystal (in order they carry an electric charge able to repulse them from one another). They also have to be large enough to be attracted by the disclination lines (for entropy reasons). In particular, they cannot just be monomer molecules that we could polymerize directly after they condense on the line. That is the reason why we make use of colloids as intermediate bricks in the materialization process of the disclination lines. These two properties of the nematics open up the opportunity to produce 3D large-scale integrated circuits, with however, a minor limitation on the connections since they have to be separated by a distance on the order of the spacing distance between substrates. The points $P$ and $Q$ that correspond to the connection points of the line to the electrodes can easily be marked by appropriate surface treatments. This gives an idea of the flexibility of the method.

Let us emphasize that an advantage of the disclination lines is that they attract the particles rather uniformly. In simpler processes that one could imagine, for instance based on electrostatic interactions, the electrodes could directly attract particles that carry an electric charge or a dipole moment. In this latter case, the particles would be attracted just as magnets attract iron filings. But, this purely electrostatic process would only yield two large aggregates stuck onto the electrodes, far from the rather uniform necklaces that we get here. Such an electrostatic process will therefore not be satisfactory.

So, a micro, and possibly nano-necklace depending on the particle size, is formed on saturating the disclination line with particles until they get in solid contact with one another, this property being denoted by a zig-zagging shape of the necklace. The necklace is then glued by means of an electropolymerization of pyrrole under appropriate conditions, mild enough to avoid gelification of the whole system, but sufficient to provide properties of mechanical cohesion and electric conduction to the wire. Clearly, the electric and mechanical properties that we get here, i.e. conductivity of the polymer film ~ 40 S.m$^{-1}$ and strength of the wire > 15 nN, respectively, are in progress compared to the previous results. More improvements are under way, essentially based on elongated particles (CNT) as the building blocks. We anticipate that both the strength and conduction of the wires will be improved and their size decreased too, provided that the CNT are attracted parallel to the disclination lines.

We gratefully acknowledge Dr. Mircea Rastei for useful enlightenings on the AFM techniques.